\documentclass[usenatbib]{mn2e}
\usepackage{graphicx,ifthen,url}
\def\ltsima{$\; \buildrel < \over \sim \;$}
\def\simlt{\lower.5ex\hbox{\ltsima}}
\def\gtsima{$\; \buildrel > \over \sim \;$}
\def\simgt{\lower.5ex\hbox{\gtsima}}
\newcommand {\uJy}{$\mu$Jy}
\newcommand {\um}{$\mu$m}

\def\um     {$\mu$m}
\def\ts     {\thinspace}
\def\kms    {\ifmmode{{\rm \ts km\ts s}^{-1}}\else{\ts km\ts s$^{-1}$}\fi}
\def\msol   {\ifmmode{{\rm M}_{\odot}}\else{M$_{\odot}$}\fi}
\def\lsol   {\ifmmode{{\rm L}_{\odot}}\else{L$_{\odot}$}\fi}
\def\zsol   {\ifmmode{{\rm Z}_{\odot}}\else{Z$_{\odot}$}\fi}

\def\ci     {\ifmmode{{\rm C}{\rm \small I}}\else{C\ts {\scriptsize I}}\fi}
\def\hi     {\ifmmode{{\rm H}{\rm \small I}}\else{H\ts {\scriptsize I}}\fi}
\def\hh     {\ifmmode{{\rm H}_2}\else{H$_2$}\fi}
\def\cone {\ifmmode{{\rm C}{\rm \small I}(^3\!P_1\!\to^3\!P_0)}
     \else{C\ts {\scriptsize I}{\small$(^3\!P_1\!\to\,^3\!P_0)$}}\fi}
\def\ctwo {\ifmmode{{\rm C}{\rm \small I}(^3\!P_2\!\to\,^3\!P_1)}
     \else{C\ts {\scriptsize I}{\small$(^3\!P_2\!\to\,^3\!P_1)$}}\fi}
\def\cij {\ifmmode{{\rm C}{\rm \small I}\,(^3P_i\to^3P_j)}\else{C\ts {\scriptsize I}\,{\small$(^3P_i\to^3P_j)$}}\fi}
\def\cii    {\ifmmode{{\rm C}{\rm \small II}}\else{C\ts {\scriptsize II}}\fi}
\def\tex {\ifmmode{{T}_{\rm ex}}\else{$T_{\rm ex}$}\fi}
\def\tmb {\ifmmode{{T}_{\rm mb}}\else{$T_{\rm mb}$}\fi}
\def\tkin {\ifmmode{{T}_{\rm kin}}\else{$T_{\rm kin}$}\fi}
\def\microns {\ifmmode{\mu{\rm m}}\else{$\mu$m}\fi}
\def\nhh   {\ifmmode{n({\rm H}_2)}\else{$n$(H$_2$)}\fi}

\newcommand{\lsun}{{\rm\,L$_\odot$}}

\loadboldmathitalic 
\title [Herschel Galaxies at z$>$1]
{Herschel-SPIRE, Far-Infrared Properties of Millimetre-Bright and
-Faint
Radio Galaxies}

\author[S.C.~Chapman et al.]
{\parbox{\textwidth}{\raggedright S.C.~Chapman,$^{1}$\thanks{E-mail: \texttt{schapman@ast.cam.ac.uk}}
R.J.~Ivison,$^{2,3}$
I.G.~Roseboom,$^{4}$
R.~Auld,$^{5}$
J.~Bock,$^{6,7}$
D.~Brisbin,$^{8}$
D.~Burgarella,$^{9}$
P.~Chanial,$^{10}$
D.L.~Clements,$^{11}$
A.~Cooray,$^{12,6}$
S.~Eales,$^{5}$
A.~Franceschini,$^{13}$
E.~Giovannoli,$^{9}$
J.~Glenn,$^{14}$
M.~Griffin,$^{5}$
A.M.J.~Mortier,$^{11}$
S.J.~Oliver,$^{4}$
A.~Omont,$^{15}$
M.J.~Page,$^{16}$
A.~Papageorgiou,$^{5}$
C.P.~Pearson,$^{17,18}$
I.~P{\'e}rez-Fournon,$^{19,20}$
M.~Pohlen,$^{5}$
J.I.~Rawlings,$^{16}$
G.~Raymond,$^{5}$
G.~Rodighiero,$^{13}$
M.~Rowan-Robinson,$^{11}$
Douglas~Scott,$^{21}$
N.~Seymour,$^{16}$
A.J.~Smith,$^{4}$
M.~Symeonidis,$^{16}$
K.E.~Tugwell,$^{16}$
M.~Vaccari,$^{13}$
J.D.~Vieira,$^{6}$
L.~Vigroux,$^{15}$
L.~Wang$^{4}$ and
G.~Wright$^{2}$}\vspace{0.4cm}\\
\parbox{\textwidth}{\raggedright $^{1}$Institute of Astronomy, University of Cambridge, Madingley Road, Cambridge CB3 0HA, UK\\
$^{2}$UK Astronomy Technology Centre, Royal Observatory, Blackford Hill, Edinburgh EH9 3HJ, UK\\
$^{3}$Institute for Astronomy, University of Edinburgh, Royal Observatory, Blackford Hill, Edinburgh EH9 3HJ, UK\\
$^{4}$Astronomy Centre, Dept. of Physics \& Astronomy, University of Sussex, Brighton BN1 9QH, UK\\
$^{5}$Cardiff School of Physics and Astronomy, Cardiff University, Queens Buildings, The Parade, Cardiff CF24 3AA, UK\\
$^{6}$California Institute of Technology, 1200 E. California Blvd., Pasadena, CA 91125, USA\\
$^{7}$Jet Propulsion Laboratory, 4800 Oak Grove Drive, Pasadena, CA 91109, USA\\
$^{8}$Space Science Building, Cornell University, Ithaca, NY, 14853-6801, USA\\
$^{9}$Laboratoire d'Astrophysique de Marseille, OAMP, Universit\'e Aix-marseille, CNRS, 38 rue Fr\'ed\'eric Joliot-Curie, 13388 Marseille
 cedex 13, France\\
$^{10}$Laboratoire AIM-Paris-Saclay, CEA/DSM/Irfu - CNRS - Universit\'e Paris Diderot, CE-Saclay, pt courrier 131, F-91191 Gif-sur-Yvette
, France\\
$^{11}$Astrophysics Group, Imperial College London, Blackett Laboratory, Prince Consort Road, London SW7 2AZ, UK\\
$^{12}$Dept. of Physics \& Astronomy, University of California, Irvine, CA 92697, USA\\
$^{13}$Dipartimento di Astronomia, Universit\`{a} di Padova, vicolo Osservatorio, 3, 35122 Padova, Italy\\
$^{14}$Dept. of Astrophysical and Planetary Sciences, CASA 389-UCB, University of Colorado, Boulder, CO 80309, USA\\
$^{15}$Institut d'Astrophysique de Paris, UMR 7095, CNRS, UPMC Univ. Paris 06, 98bis boulevard Arago, F-75014 Paris, France\\
$^{16}$Mullard Space Science Laboratory, University College London, Holmbury St. Mary, Dorking, Surrey RH5 6NT, UK\\
$^{17}$Space Science \& Technology Department, Rutherford Appleton Laboratory, Chilton, Didcot, Oxfordshire OX11 0QX, UK\\
$^{18}$Institute for Space Imaging Science, University of Lethbridge, Lethbridge, Alberta, T1K 3M4, Canada\\
$^{19}$Instituto de Astrof{\'\i}sica de Canarias (IAC), E-38200 La Laguna, Tenerife, Spain\\
$^{20}$Departamento de Astrof{\'\i}sica, Universidad de La Laguna (ULL), E-38205 La Laguna, Tenerife, Spain\\
$^{21}$Department of Physics \& Astronomy, University of British Columbia, 6224 Agricultural Road, Vancouver, BC V6T~1Z1, Canada}}

\date{\today}
\pagerange{\pageref{firstpage}--\pageref{lastpage}} \pubyear{2009}

\begin{document} 
\maketitle 
\label{firstpage}


\clearpage
\begin{abstract}
We present the first study of the far-infrared (FIR) properties of high redshift, radio-selected Ultraluminous Infrared Galaxies (ULIRGs)  using deep observations obtained with the Spectral and Photometric Imaging Receiver (SPIRE)
 from the {\it Herschel} Multi-tiered Extragalactic Survey (HerMES).
 These galaxies span a large range of 850$\mu$m fluxes from submillimetre-luminous 
$\sim$10~mJy sources ({\it SCUBA galaxies}) to $\sim$1.5~mJy from stacked SCUBA non-detections, thus likely representing a complete distribution 
of ULIRG spectral energy distributions (SEDs). 
From Keck spectroscopic surveys in the Lockman-North field we identified a sample of 31 submillimetre galaxies (SMGs) and 37 submillimetre-faint, optically-faint radio galaxies (OFRGs), all with radio-inferred IR luminosities $>10^{12}$ L$_\odot$. These galaxies were 
cross-identified    with
SPIRE 250, 350 and 500$\mu$m catalogs based on fluxes extracted at 24$\mu$m positions in the SWIRE survey, yielding a sample of
more than half of the galaxies well detected in at least two of the SPIRE bandpasses. 
By fitting greybody dust models to the SPIRE photometry together with SCUBA 
850$\mu$m measurements (for OFRGs, only $850 \mu m$ upper limits), we infer dust temperatures and far-infrared luminosities.
The OFRGs detected by SPIRE have median  $\langle$T$_{\rm dust}\rangle= 41\pm5$~K and the SMGs have
 $\langle$T$_{\rm dust}\rangle= 34\pm5$~K, both in reasonable agreement with previous (pre-{\it Herschel}) estimates,
reaffirming that the 
local FIR/radio correlation holds (at least for this subset of high-$z$ ULIRGs) at high redshift (we measure
$\langle q_{\rm IR}\rangle = 2.43\pm0.21$ using S$_{\rm IR}$ derived from greybody fit coupled with a powerlaw extrapolation to the 24$\mu$m).
Our observations firstly  confirm that a substantial fraction of OFRGs 
exhibit large infrared luminosities  corresponding to SFRs of $\sim400$ M$_\odot$/yr.
The SPIRE observations secondly confirm the higher dust temperatures for these OFRGs than similarly selected SMGs, consistent with early predictions of the submm-faint radio populations.
Our observations also clearly confirm the large infrared luminosities of most SMGs selected with S$_{850 \mu m}>5$\,mJy and radio and strong 24$\mu$m detections, corresponding to SFRs of $\sim700$ M$_\odot$/yr.
\end{abstract}

\section{Introduction}\label{s:introduction}


Submillimetre Galaxies (SMGs -- Smail, Ivison \& Blain 1997) contribute significantly to the rapid
buildup of stellar mass in the Universe at $z\,\sim\,$2.  
SMGs have a typical redshift of $z\sim2.2$ (Chapman et al.\ 2005; Wardlow et al.\ 2010), 
are massive systems ($M_* \sim 10^{10-11} M_\odot$, Swinbank et al.\ 2004; Greve et al.\ 2005; Tacconi et al.\ 2006) and are diverse in the extent and dynamics of their molecular gas reservoirs
(e.g., Tacconi et al.\ 2008; Bothwell et al.\ 2010; Ivison et al.\ 2010a,2010b). 
Interferometric observations of SMGs' CO molecular gas suggest 
that the most luminous SMGs S(850$\mu$m) $>$ 5 mJy are merging sytems (Engel et al.\ 2010) 
with high star formation efficiencies compared to typical galaxies of similar mass (Daddi et al.\
2008). 
However,
their selection at 850\um\ is inherently biased towards colder-dust
ULIRGs, particularly at z$>$1
\citep{eales00,Blain04a}. 
 In particular, since 
submm observations probe the blackbody emission of dust in the Rayleigh-Jeans regime, they are 
anti-correlated with dust temperature (S$_{850} \propto $ T$_{\rm dust}^{-4.5}$)\footnote{since $L \propto M_d T^{5.5}$ while the Rayleigh-Jeans flux $S_{RJ} \propto M_{d} T$.} for a given infrared luminosity, 
and galaxies with warmer dust can fall below the detection limit of current submm instruments.
 Recent work
\citep[e.g.][]{C04a,casey09a} has demonstrated that 850\um-faint,
high-redshift ULIRGs exist and may contribute significantly to the
cosmic star formation rate density at its peak. 
These Optically Faint-Radio 
Galaxies (OFRGs) are defined as radio sources having inferred ULIRG luminosities, with starburst (SB) or hybrid SB-AGN spectral features in the UV, and having 2.5-$\sigma$ limits on their submm fluxes which are
consistent with them being fainter than 5\,mJy at 850\,$\mu$m. They have a comoving volume density (i.e., $\sim10^{-5} Mpc^{-3}$ at $1 < z < 3$, Chapman 
et al.\ 2001, 2004), stellar masses and radio sizes comparable to SMGs, and some have a dust temperature of 
$\sim$52 K (Casey et al.\ 2009, 2010a, 2010b). 
Studies of other infrared-luminous
galaxy populations both pre-{\it Herschel} \citep[see][]{dey08,younger09,bussmann09} and post-{\it Herschel}
\citep[see][]{Oliver10,Magdis10,Roseboom10,Chanial10}
present even more evidence for diverse populations of luminous, dusty
starbursts at z\simgt1.

However, sparse
infrared data, particularly in the 50-500\um\ wavelength range
have limited the interpretation of the SMGs and OFRGs. 
Many of their fundamental properties still rely on indirect measurements. 
Direct determinations of SMG and OFRG dust temperatures are limited and have 
only been done using either a single rest-farIR point  (SHARC-II/CSO observed 350$\mu$m -- Kovacs et al.\ 2006,2010; Coppin et al.\ 2008) or with the most luminous examples of the population using the highly confused BLAST beam (250$\mu$m, 350$\mu$m \& 500$\mu$m Ivison et al.\ 2010c, Dunlop et al.\ 2010, Chapin et al.\ 2010, Casey et al.\ 2010b).

Observations of the rest-frame far-infrared emission of high-redshift galaxies with {\it Herschel} (Pilbratt et al.\ 2010) 
provide for the first time direct measurements of typical SMGs and OFRGs total infrared luminosities and dust temperatures\footnote{Although for the hotter and lower redshift examples in our sample, PACS observations on the Wien side of the spectrum are required to truly constrain the dust temperature}, which are still highly debated (e.g., Swinbank et al.\ 2008; Dave et al.\ 2008). 
Importantly, they allow for tests of earlier measurements based on extrapolation from wavelengths either shortward or longerward of the farIR peak.
Theoretical simulations of galaxy evolution continue to have difficulties in accounting for the 
inferred luminosities/star formation rates and number counts (Baugh et al.\ 2005; Dave et al.\
2009; Narayanan et al.\ 2009). Open questions remain as to whether these luminosities have been overestimated or whether for instance the IMF 
is signiÞcantly more top heavy than in the local Universe (Baugh et al.\ 2005). 
In this paper, we assess the  {\it Herschel}-SPIRE properties of these two ULIRG
populations selected using both submillimetre (submm) and radio 
wavelengths. 
Calculations assume a flat, $\Lambda$CDM cosmology with
$\Omega_\Lambda=0.7$ and $H_0=71$\,km\,s$^{-1}$\,Mpc$^{-1}$.

\begin{figure}
\centering
\includegraphics[width=0.99\columnwidth]{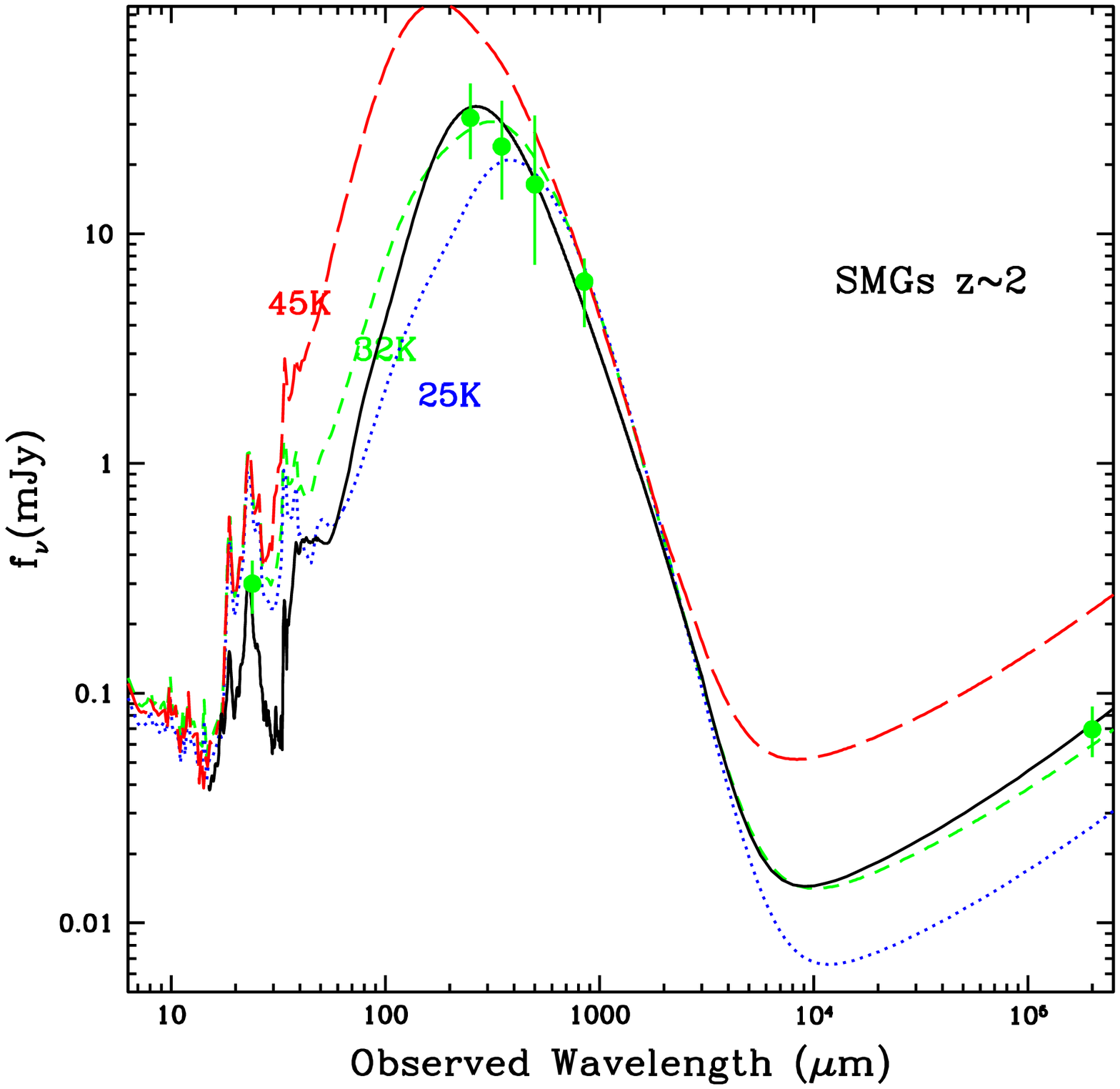}
\includegraphics[width=0.99\columnwidth]{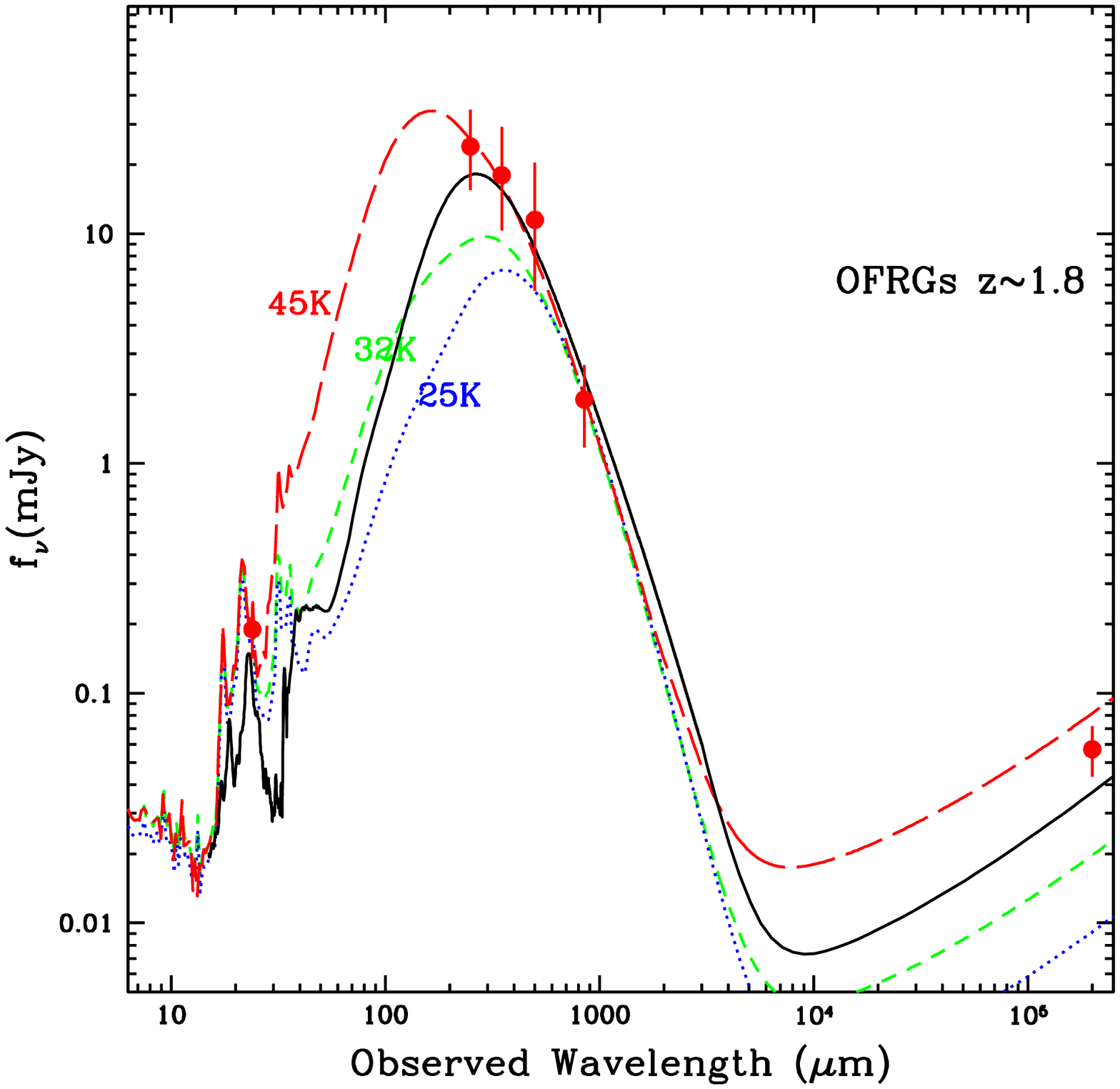}
\caption{ Average SEDs of SMGs (top) and OFRGs (bottom), after redshifting each template fit to the rest-frame, and interpolating observed data points at the average redshift of the sample. 
Circles present our SPIRE measurements, SCUBA 850$\mu$m and VLA 1.4GHz data.
Error bars show the dispersion in the sample, which is comparable to the individual flux errors combined with confusion noise.
Red, green, and blue lines (segment, dashed, dotted) present the Dale \& Helou SED templates from 25-45K.
An average SMG template from Pope et al.\ (2008) is also overlaid as solid black line.
}
\label{fig:zdist}
\end{figure}

\section{Sample \& Observations}\label{s:observations}

The HerMES\footnote{hermes.sussex.ac.uk,  Oliver et al.\ 2010 (in prep.)}
Science Demonstration Phase (SDP) observations are detailed 
in Table~1 of Oliver et al.\ (2010). 
We use deep Spectral and Photometric Imaging Receiver (SPIRE) 
(Griffin et al.\ 2010)  250, 350, 500$\mu$m  observations of SMGs and OFRGs lying in the Lockman-East (10h 51m, +57) region (a subset
of the larger Lockman-SWIRE field observations) obtained as part of this guaranteed 
time key program in a key survey field having exquisite multi-wavelength ancilliary data. 
The radio/submm sources lie within a sub-field covered by a deep VLA 1.4GHz radio pointing, within a region of 15$'\times$ 15$'$ (Ibar et al.\ 2009).
Flux densities are measured using a PSF fitting technique based on cross-identifed (XID) prior sources positions detected at 24$\mu$m (Roseboom et al.\ 2010), which accounts for blending of SPIRE source fluxes by simultaneous fitting to multiple sources. 
All sources in our fields are essentially point sources at  the resolution of {\it Herschel}.
The association between the {\it Spitzer}-IRAC \& MIPS and SPIRE 
sources is thus facilitated by the use of these priors.
 The quality of the  SPIRE  250, 350, 500$\mu$m 
catalogs has been estimated using  Monte Carlo simulations (photometry error, completeness and contamination as a function of the flux density).
In the Lockman-East field our observations have average 1$\sigma$ XID source extraction errors of 3.6 mJy, 4.2mJy  and 4.8mJy at 250, 350, 500$\mu$m respectively, and
the confusion noise is quoted as 5.8, 6.3 and
6.8mJy/beam rms  respectively by Nguyen et al.\ (2010), which we apply in quadrature to the flux errors.
 

Some SMGs and OFRGs have previously been studied spectroscopically in Lockman-east in several works  (respectively, Chapman et al.\ 2003a, 2005;  Ivison et al.\ 2005 and Chapman et al.\ 2004; Casey et al.\ 2009,2010).
The parent sample for this study consists of a sample of 376 radio sources with spectroscopic followup in the Lockman-east field. For this initial foray into the SPIRE properties of luminous, high-$z$ radio sources, we 
impose a $>5\sigma$ 24$\mu$m detection threshold (from the SWIRE catalogue --  Lonsdale et al.\ 2004) in order to extract the SPIRE fluxes at that position.
This removes some of the confusion issues arising in the full radio source sample, although it brings some bias to the study.
We adopt a catalogue based on a 24$\mu$m prior, as opposed to a radio prior, as we have
uniform 24$\mu$m coverage across these areas and a clear understanding of 
the effect of this input list on the data.
Future contributions will included a broader radio prior catalog to further explore these populations.
This catalog consists of 140 sources. 
We further restrict this sample to those with radio luminosities $>4.5\times10^{30}$ ergs/s/Hz, corresponding to an equivalent ULIRG far-IR luminosity ($>10^{12}$L$_\odot$) assuming the radio-farIR correlation (e.g.\ Condon et al.\ 1991), resulting in a sample of 68 sources.

The final sample consists of 31 securely identiÞed 
SMGs and 37 OFRGs with IR luminosities $>10^{12}$ L$_\odot$, and which are cross-matched with our 
SPIRE 250, 350, 500$\mu$m  multi-wavelength catalogs based on sources extraction using prior detections at 24$\mu$m.
 19 of the SMGs and 21 of the OFRGs in our sample are detected in at least two of the SPIRE bandpasses, 
 and form the sample we focus on in this work. 
The SMGs have an average (deboosted using the Coppin et al.\ 2006 counts and method) S$_{850\mu m}$=6mJy while the OFRGs have average S$_{850\mu m}$=1.5mJy (which we have not attempted to deboost here), together likely representing a complete distribution of ULIRG SEDs. 
In figure~1 we plot the
average SEDs of SMGs and OFRGs which is constructed by redshifting each template fit to the rest-frame, and interpolating observed data points at the average redshift of the sample. 
The dispersion in the average SED is comparable to the individual flux errors combined with confusion noise.
 Figure~1 also compares the Dale \& Helou SED templates ranging from 25-45K, and an average SMG template from Pope et al.\ (2008). Figure~3 shows the redshift distribution.

\begin{figure}
\centering
\includegraphics[width=.99\columnwidth]{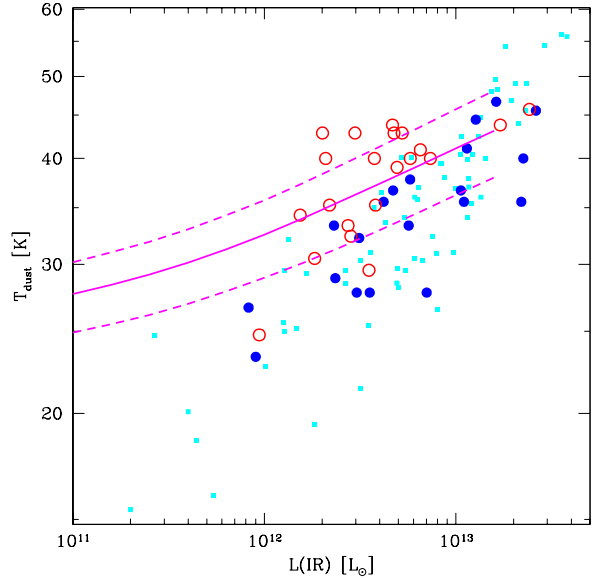}
\caption{ 
The dust temperature-luminosity relation.  
Filled blue/empty red circles are from our SMGs/OFRGs Lockman-East sample. 
Temperature uncertainties 
range from 3--7~K (depending on how well the peak is sampled) while  luminosity 
uncertainties are $\sim$0.2 dex (given the power law extrapolation down the Wien-side model fit).
The cyan squares represent results for SMGs extrapolated 
by Chapman et al.\ (2005) from radio and submm data. The Chapman et al. (2003b) derivation of the median 
and interquartile range of the T$_{\rm d}$-L$_{\rm IR}$ relation observed locally is shown by solid and dashed-dotted lines, 
linearly extrapolated to $10^{13} L_\odot$. 
}
\label{fig:firsed}
\end{figure}

\begin{figure}
\centering
\includegraphics[width=.99\columnwidth]{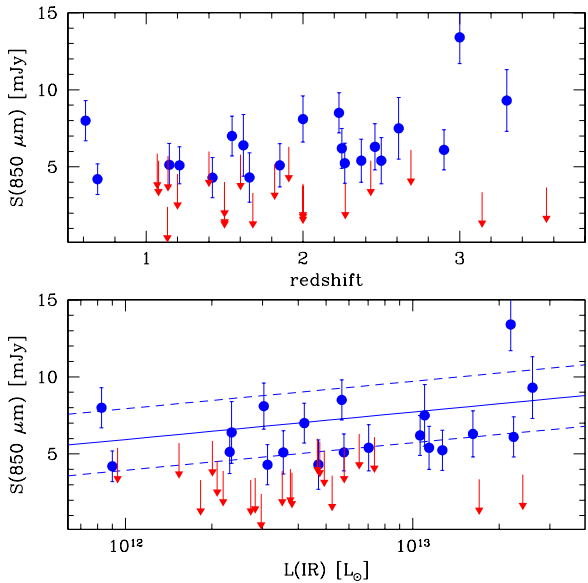}
\caption{ 
 Submm flux densities as function of the redshift and Far-Infrared luminosity, revealing both the redshift distribution of our sample and the range in 850$\mu$m fluxes probed for two samples which have similar average radio fluxes. Symbols are same as in Fig.~2 except the 
OFRGs for which we have only upper limits (flux+$2\sigma$). Solid and dashed lines show the linear fit to the S$_{850 \mu m}$ - L$_{\rm IR}$
relation for SMGs and the $1\sigma$ envelope. 
}
\label{fig:firsed}
\end{figure}

\section{Results}\label{results}

We fit the SPIRE (250\um, 350\um, and 500\um), and SCUBA (850\um) (only the upper limit in the case of the OFRGs)
flux densities to FIR dust models (see Figures 1 \& 2).
The model assumes a modified blackbody emission curve, in the optically thin approximation, with a
single dust temperature:
\begin{equation}
S_{\nu}\,\propto\,\frac{\nu^{3+\beta}}{\exp(h\nu/kT_{dust})-1}
\label{eq:blackbody}
\end{equation}
where $S_{\nu}$, the flux density, is a function of rest frequency
$\nu$, the emissivity $\beta$, dust temperature T$_{\rm d}$, and FIR
luminosity L$_{\rm FIR}$ (which governs the normalization of the
function).  
Their total infrared luminosities (L$_{\rm IR} [8-1000 \mu$m]) are inferred 
from these best fit using the far-infrared luminosity deÞnition 
(L$_{\rm FIR} [40-120 \mu$m]) given by Helou 
et al. (1988) and a color-correction term (Dale et al. 2001, L$_{\rm IR} = 1.91 \times$ L$_{\rm FIR}$). 
The model fixes emissivity to $\beta\,=\,1.5$.
The model has T$_{\rm d}$ and L$_{\rm IR}$ as free parameters.  While the luminosities here are clearly an approximation to the true L$_{\rm IR}$, we leave a more detailed estimation to the availability of shorter wavelength {\it Herschel} data.
We choose to make this model rigid since the flux errors are known to be correlated which reduces the effective number of degrees of freedom.

This single dust temperature 
characterisation provides a good description of the SMGs and OFRGs far-infrared SED, however for 7 SMGs and 9 OFRGs, this single 
dust temperature model yields high $\chi^2$ values.
All these galaxies appear 
either to be the less luminous examples,  exhibit far-infrared colours possibly suggestive of hotter systems, or may simply lie in confused regions.
However, given the extreme dust masses ($\sim10^9$ M$_\odot$) of SMGs, one should 
expect optical depths to matter, especially in the shorter wavelength SPIRE 
bands. Assuming typical emission scales of $\sim$2~kpc, we expect optical
depths of 1--2 in the 250$\mu$m and 350$\mu$m bands for a $z\sim2$ SMG (see Kovacs et al.\ 2010). 
While the effect is still moderate such that the optically 
thin approximation is crudely correct, the high $\chi^2$ found for some of the galaxies may be an
 indication of the shortcoming of the single-T, optically thin model.

Our results appear to be  relatively insensitive to $\beta$ and to the single dust component 
model; with $\beta$ = 2, the differences in
T$_{\rm d}$ are $\Delta$T$_{\rm d} \pm3$ K. 
In addition, the multiple dust components model of Dale \& Helou (2002) yields rest-frame infrared colors 
(S$_{60\mu{\rm m}}/$S$_{100\mu{\rm m}}$), 
or equivalently T$_{\rm d}$ , which are in reasonable agreement with those inferred from 
our greybody analysis. 
In figure~1, we also overplot a composite SMG spectrum, from Pope et al.\ (2008), 
normalized to 24\um\ flux density.  

The T$_{\rm d}$-L$_{\rm IR}$ plane forms an important diagnostic for assessing the properties of luminous IR galaxies and their spectral shapes.
In Figure 2,  the locations of our SMGs and OFRGs on the T$_{\rm d}$-L$_{\rm IR}$  plane are shown.
These 
estimates are amongst the first direct observational measurements of the dust temperatures and the infrared 
luminosities of OFRGs. Previous studies with SHARC-II and BLAST have attempted similar analyses mainly on SMGs with caveats presented in the Introduction (Kovacs et al.\ 2006, 2010, Coppin et al.\ 2008; Ivison et al.\ 2010c, Dunlop et al.\ 2010, Chapin et al.\ 2010, Casey et al.\ 2010, Amblard et al.\ 2010).
Our observations suggest that high redshift ULIRGs show a wide range 
of dust temperatures. 
At  intermediate infrared luminosities of ULIRGs (L$_{\rm IR}$ $\sim 3-5\times10^{12}$ L$_\odot$), the dust 
temperature dispersion observed in our sample might suggest a higher T$_{\rm d}$-L$_{\rm IR}$ scatter than deduced from local
IRAS galaxies  
by Chapman et al.\ (2003) and Chapin et al.\ (2009).
OFRGs tend to be biased towards hotter dust 
temperatures, with median $\langle$T$_{\rm d}\rangle=41\pm5$~K (quoting standard deviations of the sample) and $\langle {\rm L_{IR}}\rangle=3.8\times10^{12}$ L$_\odot$, while
SMGs appear to have lower dust temperatures with median $\langle$T$_{\rm d}\rangle=34\pm5$~K and $\langle {\rm L_{IR}}\rangle=7.1\times10^{12}$ L$_\odot$.
There is a trend of increasing T$_{\rm d}$ with increasing L$_{\rm IR}$ for both SMGs and OFRGs.
Firstly the S$_{\rm 850 \mu m}$ selection 
effectively selects M$_{\rm d}\times$T$_{\rm d}$, thus for a given mass warmer galaxies are favoured, which will 
be more luminous. Secondly, we expect something of a T--L correlation because of the temperature 
bias in our luminosity estimates (see below).


In Figure 3, we depict the  
locations of SMGs and OFRGs on the S(850$\mu$m)--$z$ and --L$_{\rm IR}$ planes.
The weak correlations between S(850$\mu$m) and L$_{\rm IR}$, (suggested by Pope et al.\ 2005) are rendered meaningless when the sample of OFRGs are considered together with SMGs.
Selection effects are a concern in these diagrams, since our focus on submm-bright or submm-faint radio sources
with relatively bright 24$\mu$m detections, as well as SPIRE detections means that we may have missed populations of galaxies which would fill in and further elucidate these trends.

\section{Discussion}\label{sec:discussion}

Comparing to previous studies of SMG dust temperatures, Chapman et al.\ (2005) found  T$_{\rm d} = 36\pm7$~K by using the FIR/radio correlation (e.g., Condon et al.\ 1991). As mentioned earlier, these measurements relied on indirect indicators of T$_{\rm d}$.
In order to more directly compare our SPIRE results with previous studies, we applied the method of Chapman et al.\ (2005) to our SMG sample,  using only the radio and 
850$\mu$m data and fitting Dale \& Helou (2002) dust SED templates.
T$_{\rm d}$ is then derived by applying a mapping from their S$_{60\mu{\rm m}}/$S$_{100\mu{\rm m}}$. 
Using the full SED information for these SPIRE sources, we find that
T$_{\rm d}$, measured from $S_{850}$ and $L_{1.4GHz}$ alone,
overestimates the dust temperature by 2$\pm$2\,K for SMGs, and on average overestimates
the dust temperature of OFRGs by 7$\pm$4\,K.
This is remarkable agreement, given the scatter in the IR-radio relation and poorly defined 850$\mu$m fluxes for individual OFRGs.
Kovacs et al.\ (2006) observed 21 SMGs at 350\um\ and measured FIR
luminosities independent of radio luminosity using FIR data ranging
from observed 350\um\ to 1200\um.  They found that the local FIR/radio
correlation overestimates FIR luminosity by factors of
$\sim$0.2-0.4\,dex for SMGs, which agrees with our SPIRE analysis.

To demonstrate these discrepancies more clearly, we estimate the
bolometric $q_{\rm IR}$, the ratio of IR integrated flux to radio flux, as
described in Ivison et al.\ (2010d):
\begin{equation}
q_{\rm IR}\,=\,\log(\frac{S_{IR}}{3.75\times10^{12}W\,m^{-2}})\,-\,\log(\frac{S_{1.4GHz}}{W\,m^{-2}\,Hz^{-1}})
\end{equation}
where S$_{\rm IR}$ taken from rest frame 8 to 1000\um\ as described before and the radio
flux density $S_{1.4GHz}$ is K-corrected, assuming a radio
slope, $\alpha\,=\,0.75$ (Ibar et al.\ 2010). 
We note,
however, that the choice of radio spectral slope only affects the
$q_{\rm IR}$ ratio by $<$10\%.
We also note that the Dale \& Helou SED templates assume a FIR/radio correlation with an effective
$\langle q_{\rm IR}\rangle$= 2.62 (Yun et al. 2001)\footnote{The Helou et al.\ (1988) 
definition of `q' using L[40-120$\mu$m] differs from  our adopted Ivision et al.\ (2010d) definition of q$_{\rm IR}$  
with L[8-1000$\mu$m], all other terms being the same. Thus with an average colour 
conversion of 1.91, we find q$_{\rm IR} \sim$ q $+ 0.28$.}.
%
Since our S$_{\rm IR}$ estimates are based on the greybody fits, we extrapolate from the peak to the 24$\mu$m point
with an appropriate power-law, yielding $\langle q_{\rm IR}\rangle$= 2.43$\pm0.2$,  
thereby suggesting a reliable estimate of the infrared luminosity 
of $z\sim2$ ULIRGs can be obtained from their radio luminosity using a similar $q$ to that inferred locally, 
as inferred in Kovacs et al.\ (2006,2010).

The SMG composite template (Pope et al.\ 2008) used in Fig.~1 is derived based on mid-IR to FIR data of
SMGs to date. While it reasonably fits the SPIRE FIR data of the SMG sample,
in some cases, it under/overestimates the FIR luminosities by
$\pm$0.5\,dex, and unsurprisingly is much worse at describing the OFRG sample.  
This illustrates how a 24\um-normalized SED fitting
procedure, which is common in the literature (Desai et al.\ 2009)
places poor constraints on the breadth of
FIR properties of ULIRG samples, especially in the absence of direct
FIR measurements.
Similarly, the use of the 24$\mu$m emission and of the Chary \& Elbaz SED library yields an inaccurate estimate of the infrared luminosity characterized by a large scatter 
($\sim 0.5$ dex) and a systematic overestimation ($\sim \times$2 times) in the luminosities of ULIRGs.
For radio-ULIRGs
under investigation here, luminosity extrapolations based on the radio emission are considerably more reliable 
than those based on the mid-infrared emission (e.g., Elbaz et al.\ 2010, although they have shown that 24$\mu$m is a good representation of L$_{\rm IR}$ for lower redshift galaxies). 

It is of interest to estimate the expected  infrared luminosities of our SPIRE-undetected SMGs and OFRGs, although in the latter case there is some likelihood that an AGN is boosting the radio luminosity. 
 The 
SPIRE nondetections of SMGs are generally found to be consistent with the properties inferred from the detections.
We use $\langle q_{\rm IR}\rangle$=2.43 and apply the Dale \& Helou SED templates normalized to these infrared luminosities.
For the SMGs we find that for 5 out of 11 undetected SMGs, SPIRE flux densities 
inferred using these fits lie below the detection threshold. Of the 6 sources 
with  predictions for SPIRE fluxes above the detection threshold, 3 
have evidence in the UV spectra for AGN contributions.
For the OFRGs the lack of 850$\mu$m detection and SPIRE detections makes it likely that an AGN is generating the radio luminosity.

The main concern with this study is just how representative the sample of ULIRGS is to  
the wider SMG and OFRG populations given the use of a 24$\mu$m prior.
The 24$\mu$m detection could favour those galaxies that are either warmer, or have 
larger fractions of warm (T$>$35K) gas than average, biasing both the luminosities and temperatures. 
Apart from misrepresenting 
the class of objects as a whole, the different temperature distributions also 
question the usability of local templates (which presumably have a more average
temperature distribution). Also, because the 24$\mu$m band falls into a range of the rest-frame 
spectrum where there are a lot of aromatic features, it is even more 
complicated to assess what  subclasses are selected at what redshift.
While our study has 
some clear advantages compared to prior FIR studies of SMGs and OFRGs, the additional 
bias is a disadvantage, motivating a more thorough investigation of the SPIRE properties of more complete samples.

Our observations firstly  confirm that a substantial fraction of OFRGs ($\sim$50\% of the 24$\mu$m XID SPIRE sample) exhibit large infrared luminosities  corresponding to SFRs of 380 M$_\odot$/yr
(SFR [M$_\odot$/yr] = 1$\times 10^{-10}$ L$_{IR}$ [L$_\odot$],  assuming a Chabrier IMF). 
The complicated selection of our sample in this study, including the required 24$\mu$m XID prior for SPIRE flux extraction, means that a precise volume density for these IR-luminous OFRGs is not trivial to estimate. 
As a lower limit, given the OFRGs without 24$\mu$m XID prior which have yet to be studied with {\it Herschel},
the IR-luminous OFRGs are about 1/4 as numerous as equivalent luminosity SMGs at $z\sim2$, or $>7\times10^{-6}$~Mpc$^{-3}$ for L$_{\rm IR}=2-6\times10^{12}$~L$_\odot$.
The SPIRE observations secondly confirm the higher dust temperatures for these OFRGs than similarly selected SMGs with SPIRE observations, consistent with our early predictions of the submm-faint radio populations.
Understanding the different dust temperatures and the  connections to  SMGs is possible if we take the 
luminosities of SMGs and OFRGs as dominated by star-formation, even if an AGN contribution cannot be ruled out.
Since L $\propto$ M$_{\rm d}$ T$^{5.5}$, warmer dust implies less dust. As such, 
given the similar SMG and OFRG luminosities but their different typical 
temperatures, a possible conclusion is simply that OFRGs have about a third of the 
dust content relative to stellar mass. 
This would naturally suggest that OFRGs 
are simply a different phase of the starburst than the cooler SMGs. Either 
OFRGs are an earlier phase, before the full-scale build-up of dust mass, or a 
later phase, where much of the dust is blown out by superwinds or consumed by a central black hole.

Our observations also clearly confirm the remarkably large infrared luminosities of most SMGs (S$_{850 \mu m}>$ 5 mJy) which correspond to SFRs of $710$ M$_\odot$/yr.
The remaining 12 SMGs in our Lockman-East catalog which are not
found in the XID SPIRE catalog do not distinguish themselves in terms of 850$\mu$m or 24$\mu$m fluxes. Although their radio-850$\mu$m inferred T$_{\rm d}$ is cooler on average than those presented here, it is likely source confusion in the SPIRE maps which limits their study. 
More detailed evolutionary understanding of high-$z$ ULIRGs will be facilitated by future
studies of   deep {\it Herschel} observations and including the shorter wavelength bands.

\section*{Acknowledgments}
We thank the referee, Attila Kovacs, for a careful reading and constructive criticism of the manuscript.
The data presented in this paper will be 
released through the {\it Herschel} Database in Marseille HeDaM 
(hedam.oamp.fr/HerMES). SPIRE has been developed by a consortium of institutes led by Cardiff Univ. (UK) and including 
Univ. Lethbridge (Canada); NAOC (China); CEA, LAM (France); 
IFSI, Univ. Padua (Italy); IAC (Spain); Stockholm Observatory 
(Sweden); Imperial College London, RAL, UCL-MSSL, UKATC, 
Univ. Sussex (UK); Caltech, JPL, NHSC, Univ. Colorado (USA). 
This development has been supported by national funding agencies: 
CSA (Canada); NAOC (China); CEA, CNES, CNRS (France); ASI 
(Italy); MCINN (Spain); SNSB (Sweden); STFC (UK); and NASA 
(USA). Data presented herein were obtained using the W.\ M.\ Keck
Observatory, which is operated as a scientific partnership among
Caltech, the University of California and NASA. The Observatory was
made possible by the generous financial support of the W.\ M.\ Keck
Foundation.


\begin{thebibliography}{}

\bibitem[]{} Amblard, A., Cooray, A., Serra, P., Temi, P., Barton, E., Negrello,  M., Auld, R., Baes, M., et al., 2010, 
A\&A, aph/1005.2412

\bibitem[]{} Baugh, C. M., Lacey, C. G., Frenk, C. S., et al. 2005, MNRAS, 356, 1191

\bibitem[Blain et al.\ 2004]{Blain04a} Blain, A.W., Chapman, S.C., Smail, I., Ivison R.,
2004, ApJ, 611, 52 

\bibitem[Bussmann et al.\ 2009]{bussmann09}
Bussmann R. S., et al., 2009, ApJ, 705, 184

\bibitem[Casey et al.\ 2009]{casey09a} Casey, C. M., Chapman, S. C., Beswick, R. J., et al. 2009, MNRAS, 399, 121 

\bibitem[Casey et al.\ 2010]{casey10a} Casey, C. M., Chapman, S. C., Neri, R., et al. 2010a, MNRAS, arXiv0910.5756 

\bibitem[Casey et al.\ 2010]{casey10b} Casey, C. M.,  Chapman, S., Smail, I., et al., 2010b, MNRAS, in press

\bibitem[Chanial et al.\ 2010]{Chanial10} Chanial, P.\ et al.\ 2010, MNRAS, in prep

\bibitem[]{Chapman01} Chapman, S.\ C., Richards, E.\ A., Lewis, G.\ F.,
Wilson, G., Barger, A.\ J., 2001, ApJ, 548, L147



\bibitem[Chapman et al.\ 2003]{C03} Chapman, S.\ C., Blain, A., Ivison, R., Smail, I.,
      2003a, Nature, 422, 695 

\bibitem[]{CHLD03} Chapman, S.\ C., Helou, G., Lewis, G.\ F., Dale, D.,
      2003b, ApJ, 588, 186

\bibitem[]{} Chapman, S. C., Smail, I., Blain, A. W., Ivison, R. J. 2004, ApJ, 614, 671

\bibitem[]{} Chapman, S. C., Blain, A. W., Smail, I., Ivison, R.\ 2005, ApJ, 622, 772

\bibitem[Chapin et al.\ 2009]{Chapin09} Chapin, E., Hughes, D., Aretxaga, I.,  2009, MNRAS, 393, 653	

\bibitem[Chapin et al.\ 2010]{Chapin10}  Chapin, E., et al., 2010 arXiv1003.2647C

\bibitem[Chapman et al.\ 2005]{C05} Chapman, S.\ C., Blain, A., Smail, I., Ivison, R.\ J.,
      2005, ApJ, 622, 344

\bibitem[Chapman et al.\ 2004a]{C04a} Chapman, S.\ C., Smail, I., Blain, A., Ivison, R.\ J.,
      2004, ApJ,  614, 67
      

\bibitem[]{Condon} Condon, J.\ J., Anderson, M.\ L., Helou, G., 1991,
ApJ, 376, 95

\bibitem[]{} Coppin, K., et al.,  2006, MNRAS, 372, 1621 

\bibitem[]{} Coppin, K., et al., 2008, MNRAS, 378, 214

\bibitem[]{Daddi08} Daddi, E., Rottgering, H.\ J.\ A., Labbe, I.,
 Rudnick, G., Franx, M., Moorwood, A.\ F.\ M., Rix, H.\ W., van der
 Werf, P.\ P., van Dokkum, P.\ G., 2008, ApJ, 628, 50

\bibitem[]{dale01}Dale, D., Helou, G.,
     Contursi, A.; Silbermann, N.\ A..
     Kolhatkar, S.,  2001, ApJ, 562, 142

\bibitem[]{} Dale, D. A., Helou, G. 2002, ApJ, 576, 159

\bibitem[]{}  Dave, R., Finlator, K., Oppenheimer, B. D., et al. 2010, in press

\bibitem[Desai et al.\ 2009]{desai09}Desai V., et al., 2009, ApJ, 677, 943

\bibitem[Dey et al.\ 2008]{dey08}Dey A., et al., 2008, ApJ, 677, 943

\bibitem[Dunlop et al.\ 2010]{dunlop10}Dunlop, J., et al., 2010, arXiv0910.3642D

\bibitem[Eales et al.\ 2000]{eales00} Eales, S., Lilly, S., Webb, T., Dunne, L., Gear,
W., Clements, D., Yun, M., 2000, AJ, 120, 2244

\bibitem[]{Elbaz} Elbaz, D., et al., 2010, A\&A, in press, (arXiv:1005.2859)

\bibitem[]{Greve05} Greve, T., et al., 2005, ApJ, 621, 124

\bibitem[]{Griffin} Griffin, M.,   Abergel, A., Abreu, A., et al. 2010, A\&A, in press

\bibitem[]{} Helou, G., Khan, I. R., Malek, L., Boehmer, L. 1988, ApJS, 68, 151 

\bibitem[]{} Ibar, E., Ivison, R., Best, P., et al., 2010 MNRAS, 401L, 53

\bibitem[]{} Ibar, E., Ivison, R., Biggs, A., Lal, D., Best, P., Green, D.\ 2009, MNRAS, 397, 281

\bibitem[]{I05} Ivison, R.\ J., et al., 2005, MNRAS, 364, 1025

\bibitem[]{I10} Ivison, R.\ J., et al., 2010a, MNRAS, 404, 198  

\bibitem[]{I10} Ivison, R.\ J., et al., 2010b, MNRAS, arXiv:1009.0749 

\bibitem[]{I10} Ivison, R.\ J., et al., 2010c, MNRAS, 2010, 402, 245 

\bibitem[]{I10} Ivison, R.\ J., et al., 2010d, MNRAS, 2010, arXiv1005.1072 

\bibitem[Kovacs et al.\ 2006]{kovacs06a} Kovacs, A., et al.\ 2006, ApJ, 634, 31

\bibitem[Kovacs et al.\ 2010]{Kovacs10} Kovacs, A., et al., 2010, ApJ, 717, 29

\bibitem[]{I10} Lonsdale et al.\ 2004, ApJS, 154, 54

\bibitem[Magdis et al.\ 2010]{Magdis10} Magdis et al.\ 2010, MNRAS, submitted

\bibitem[Nguyen et al.\ 2010]{nguyen} Nguyen et al.\ 2010, MNRAS, in press

\bibitem[Oliver et al.\ 2010]{Oliver10}  Oliver, Seb; Frost, M.; Farrah, D.; Gonzalez-Solares, E.; Shupe, D. L.; Henriques, B.; Roseboom, I.; Alfonso-Luis, A. et al.\ 2010, MNRAS, in press


\bibitem[]{}Pilbratt, G. et al. 2010, A\&A, in press


\bibitem[Pope et al.\ 2008]{pope08} Pope, A., et al.\  2008, ApJ, 603, L13


\bibitem[Roseboom et al.\ 2010]{Roseboom10} Roseboom, I., et al.\ 2010, MNRAS, in press

\bibitem[Seymour et al.\ 2009]{seymour09a} Seymour, N.\ et al.\ 2009, MNRAS, 398, 1573 

\bibitem[]{} Smail, I., Ivison, R., Blain A., 1997, ApJ, 490, L5

\bibitem[]{} Wardlow, J.\ et al.\ 2010, MNRAS,  submitted

\bibitem[]{}Yun, M. S., Reddy, N. A., \& Condon,J.J., 2001, ApJ, 554, 803











\bibitem[Younger et al.\ 2009]{younger09}
Younger J. D., et al., 2009, MNRAS, 394, 1685 

\end{thebibliography}

\label{lastpage}

\end{document}